\documentclass[%
reprint,
 amssymb, amsmath,%
 aps,
 prl,
]{revtex4-1}

\usepackage[pdftex]{graphicx}
\graphicspath{{./figures/}}
\DeclareGraphicsExtensions{.pdf,.jpeg,.png}

\usepackage{stmaryrd}
\usepackage{docs}%
\usepackage{bm}%
\usepackage[colorlinks=true,linkcolor=blue]{hyperref}%
\expandafter\ifx\csname package@font\endcsname\relax\else
 \expandafter\expandafter
 \expandafter\usepackage
 \expandafter\expandafter
 \expandafter{\csname package@font\endcsname}%
\fi
\begin{document}

\title{Influence of the ESR saturation on the power sensitivity of cryogenic sapphire resonators}%

\author{Vincent Giordano}%
\email{giordano@femto-st.fr}
\author{Serge Grop, Pierre-Yves Bourgeois, Yann Kersal\'e and Enrico Rubiola}
\affiliation{FEMTO-ST Institute - UMR 6174\\
CNRS / ENSMM / UFC / UTBM\\
26 Chemin de l'\'Epitaphe\\
25000 Besan\c{c}on -- FRANCE}%
\date{2014 may, 12}%
\revised{}%

%Centre National de la Recherche Scientifique, \'Ecole Nationale Sup\'erieure de M\'ecanique et des Microtechniques, Universit\'e de Franche-Comt\'e, Universit\'e Technologique de Belfort Montb\'eliard

\begin{abstract}

Here, we study the paramagnetic ions behavior in presence of a strong 
microwave electromagnetic field sustained inside a cryogenic sapphire whispering gallery mode resonator. The high frequency measurement resolution  that can be now 
achieved by comparing two CSOs permit for the first time to observe clearly the non-linearity of the resonator power sensitivity. 
These observations that in turn allow us to optimize the CSO operation, are well explained by  the Electron Spin Resonance (ESR) saturation of 
the paramagnetic impurities contained in the sapphire crystal.

\end{abstract}

\maketitle

%\tableofcontents

\section{Introduction}
Sapphire single crystal is a key material for numbers of very demanding scientific and technological applications. 
Even in its purest form, the sapphire single crystal always contains some paramagnetic impurities, which come from the raw material or are result of contamination during the growth process \cite{barish02,benmessai13-prb}. 
Despite their low concentration these accidental dopants turn out to be very useful in some innovative systems intended for high resolution measurements.  
The Cryogenic Sapphire Oscillator (CSO) incorporating a cryogenic sapphire whispering gallery mode resonator is currently the most stable frequency source. It achieves in an autonomous and reliable version
a relative frequency stability better than $1\times 10^{-15}$ for integration times $\tau \leq 10,000$ s, with at long term a frequency drift of $2\times 10^{-15}/$day \cite{uffc-2011-long-term-stab,RSI-2012}. 
The thermal compensation induced by the accidental paramagnetic impurities that substitute to Al$^{3+}$ is essential for the achievement of the highest frequency stability \cite{mann92-jpd,luiten96}. 
A high Q-factor microwave cryogenic resonator containing a small amount of Fe$^{3+}$ can also be used as the amplifying medium to design a zero-field solid-state 12 GHz Maser \cite{apl05-maser,prl-2008} or as the non-linear element for microwave third harmonic generation \cite {creedon12-prl-four-wave-mix, creedon12-prl-freq-conv}
Paramagnetic impurities in sapphire are also good candidates for the realization of quantum electrodynamics measurements\cite{farr13}.\\

In this paper we study the behavior of paramagnetic dopants diluted in a sapphire high-Q microwave resonator in which a strong electromagnetic field is sustained. The Van-Vleck model is applied to calculate the dc-magnetic susceptibility induced at low power by the different paramagnetic species that can be found in the high quality sapphire crystal. The interaction between the paramagnetic dopants and the RF electromagnetic field is described through the classical two-levels Bloch equations. We show how the saturation of the Electron Spin Resonances (ESR) leads to a non-linear power sensitivity of the resonator frequency. Eventually the stabilisation of the injected power to the resonator turnover point allow us to optimise the oscillator frequency stability, which reaches a flicker floor of $3\times 10^{-16}$ extending until 10,000 s.

%%%%%%%%%%%%%%%%%%%%%%%%%
%%%%%%%%%%%%%%%%%%%%%%%%%
\section{Paramagnetic ions description}
%%%%%%%%%%%%%%%%%%%%%%%%%
%%%%%%%%%%%%%%%%%%%%%%%%%

%%%%%%%%%%%%%%%%%%%%%%%%%
\subsection{Standard magnetic susceptibility model}
%%%%%%%%%%%%%%%%%%%%%%%%%

The presence of paramagnetic dopants in the crystal matrice and its impact on the propagation of an electromagnetic wave at the frequency $\nu$  is accounted for through the magnetic susceptibility $\chi(\nu)=\chi'(\nu)+j\chi''(\nu)$. $\chi'$ and $\chi''$ represent respectively  the phase shift and the power absorption induced by the Electron Spin Resonance (ESR) \cite{siegman_maser}. Let us assume that the sapphire crystal contains a density $N$ of a paramagnetic ion presenting in its ground state two energy levels $ |m\rangle$ and $ |n\rangle$ separated by the ion ESR frequency $\nu_{mn}$.  Solving the Bloch equations for such a two levels system interacting with the wave at the frequency $\nu$ leads for the real part of the susceptibility to a dispersive lonrentzian function that nulls at $\nu_{mn}$ \cite{vanier-audoin-T1}:
\begin{equation}
\chi'(\nu)= \chi_{0} \dfrac{(2\pi\tau_{2})^{2}(\nu-\nu_{mn}) \nu_{mn}}{1+(2\pi\tau_{2})^{2}(\nu-\nu_{mn})^{2}+\Omega^{2}\tau_{1}\tau_{2}}
\label{equ:chiprim}
\end{equation}
where $\tau_{1}$ and $\tau_{2}$ are the ion spin-lattice and spin-spin relaxation times, respectively. $\Omega$ is the Rabi frequency and $\chi_{0}$ is the dc-susceptibility. Similarly, we found an absorption lorentzian lineshape for $\chi''(\nu)$, whose linewidth at low power is $\Delta \nu_{mn}= {1/\pi \tau_{2}}$. \\

$\Omega$ is proportional to the ESR transition probability $\sigma_{mn}$ and to the RF magnetic field $B_{RF}$:
\begin{equation}
\Omega= \dfrac{g \mu_{B} B_{RF}}{\hbar}\sigma_{mn}
\label{equ:omega}
\end{equation}
where $g$ is the Land\'e factor, $\mu_{B}$ the Bohr Magneton and $\hbar$ the reduced Planck constant.
The term $\Omega^{2}\tau_{1}\tau_{2}$ in the denominator of the equation \ref{equ:chiprim} is the saturation parameter proportional to the electromagnetic power felt by the ions. This term is generally neglected in the analysis of the paramagnetic ion behavior in the  CSO sapphire resonator. However, as we will see later, when the electromagnetic wave is confinded in a high-Q factor cryogenic sapphire resonator the saturation of the ESR can no longer be neglected. 

The dc-susceptibility $\chi_{0}$ results from the distribution of the ions on their energy levels through the effect of the thermal agitation. The derivation of $\chi_{0}$ at a given absolute temperature $T$ is straightforward by assuming i) the energy separation between the ion ground state and excited state 
is large compared to $k_{B}T$, ii) the ion orbital momentum is totally quenched by the crystal field and thus the ion in the crystal lattice behaves like a free spin $S$. With these assumptions, $\chi_{0}$ follows the Curie law \cite{buschow}:

\begin{equation}
\chi_{0}^{C} = \mu_{0} N \dfrac{g^{2} \mu_{B}^{2}  }{3 k_{B} T} S(S+1)
\label{equ:chi0C}
\end{equation}
where $\mu_{0}$ is the permeability of free space and $k_{B}$ the Boltzmann constant.

%%%%%%%%%%%%%%%%%%%%%%%%%
\subsection{On the validity of the ESR description}
%%%%%%%%%%%%%%%%%%%%%%%%%

The Curie law has been derived for a free system of spin $S$, which consists in $2S+1$ levels equally spaced. Departures from the Curie law are well known 
for Sm$^{3+}$ and Eu$^{3+}$ as for these two rare-earth ions  the first excited multiplet populations can not be neglected.
Even for less exotic ions, in a real crystal the ground state is splitted by the crystal field in multiple degenerated Kramer's doublets separated by a 
Zero Field Splitting (ZFS)\cite{majlis}. The table \ref{tab:ions} gives the characteristics of the dominant paramagnetic species that can be found in high purity sapphire crystals.

\begin{table}[h!!!!!]
\centering
\caption{ESR of the Iron-group paramagnetic ions that can be found in high-purity sapphire crystals.}
\label{tab:ions}
\begin{tabular}{lccrlrrr}
\hline
Ion			& $S$	&	Ground state	& $\nu_{mn}$~~	& $\sigma_{mn}$& $\tau_{1}$	& $\tau_{2}$ & Ref.	\\
			&		&        transitions	& (GHz)		&			&		(ms)	&	(ns)	     &	\\	
\hline
\hline
Cr$^{3+}$  	&3/2		&$|\frac{1}{2}\rangle \shortrightarrow |\frac{3}{2}\rangle$& 11.4~~	 	& 1.00			 & 200 & 7&\cite{standley65,siegman_maser} \\
\\
Mo$^{3+}$  	& 3/2	&$|\frac{1}{2}\rangle \shortrightarrow |\frac{3}{2}\rangle$& 165.0~~		& 2.00 			& 0.1 & 12&\cite{sharoyan74,kocharyan79} \\
\\
Fe$^{3+}$  	& 5/2	&$|\frac{1}{2}\rangle \shortrightarrow |\frac{3}{2}\rangle$& 12.0~~ 		& 2.00			& 10  & 20 &\cite{benmessai13-prb,bogle59} \\
			&		& $|\frac{3}{2}\rangle \shortrightarrow |\frac{5}{2}\rangle$	& 19.3~~	& 1.25			& 	&  &\\
			&		& $|\frac{1}{2}\rangle \shortrightarrow |\frac{5}{2}\rangle$	& 31.3~~	& 0.0024			& 	&  &\\
\\			
\hline
\end{tabular}
\end{table}

\begin{figure}[h!!]
\centering
\includegraphics[width=\columnwidth]{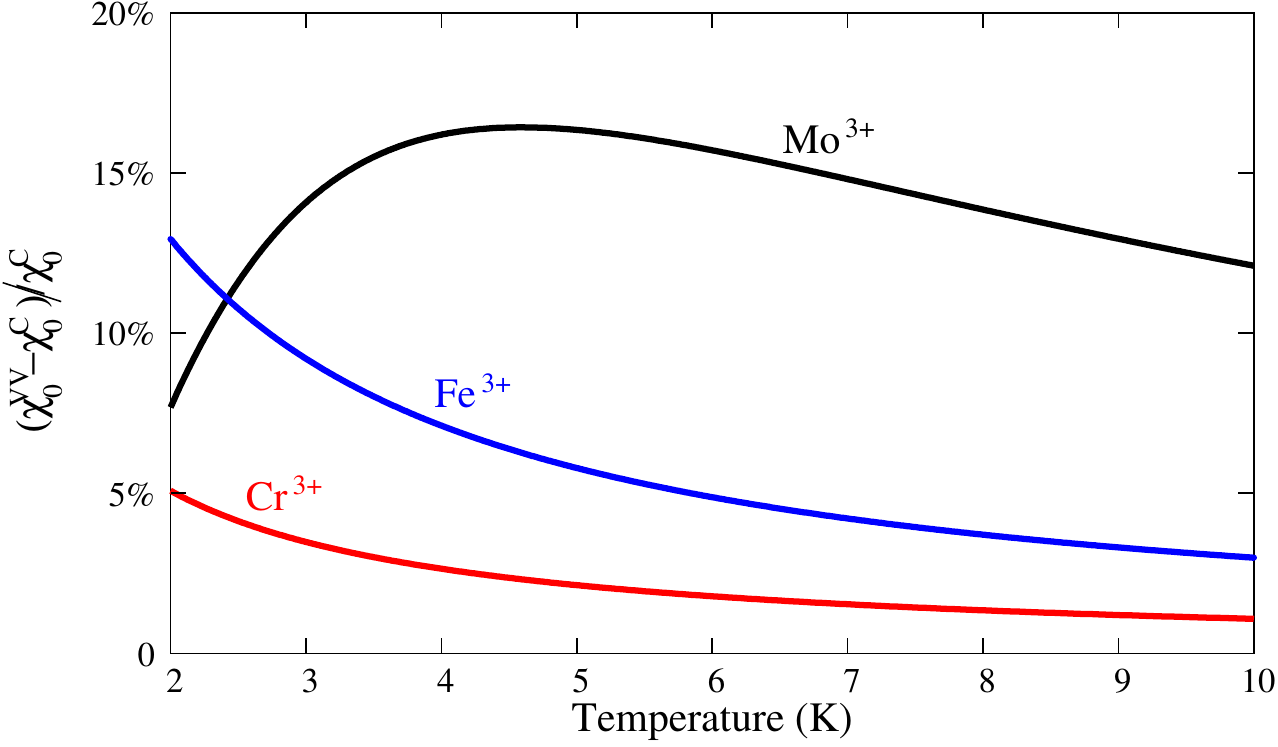}
\caption{Comparison between the Curie and the Van Vleck laws : $(\chi_{0}^{VV}-\chi_{0}^{C})/\chi_{0}^{C}$ for Cr$^{3+}$ (red), Fe$^{3+}$ (blue) and Mo$^{3+}$ (black) in Al$_{2}$O$_{3}$ .}
\label{fig:comparison-VV-C}
\end{figure}

Strictly speaking, for these ions the dc-susceptibility should be calculated by using the Van Vleck equation \cite{vanvleck78}:
\begin{equation}
\chi_{0}^{VV} = \mu_{0} N \dfrac{\sum\limits_{m}({E_{m}^{(1)}}^{2}-2E_{m}^{(2)})\exp(-E_{m}^{(0)}/k_{B}T )}{\sum\limits_{m} \exp(-E_{m}^{(0)}/k_{B}T )} 
\label{equ:chi0VV}
\end{equation}
providing we know a Taylor expansion as a function of the applied magnetic field $B_{0}$ of each populated energy levels,
i.e. $E_{m} = E_{m}^{(0)} + E_{m}^{(1)}B_{0}+ E_{m}^{(2)}B_{0}^{2}+\mathcal{O}^{3}$. In a number of situations, the cumbersome calculation of the Van Vleck 
coefficients $E_{m}^{(k)}$ is avoided and the dc-susceptibitily is assumed to follow the Curie law. However the current measurement resolution is such as it is
necessary to know the degree of validity of this assumption. In \cite{boca04}, R. Bo\u{c}a uses the spin-hamiltonian formalism to derive the Van Vleck coefficients, 
and thus the dc-susceptibility expression for various systems presenting a ZFS. From his results we calculated  $\chi_{0}^{VV}$ for Cr$^{3+}$, Fe$^{3+}$ 
and Mo$^{3+}$ in Al$_{2}$O$_{3}$ neglecting the rhombic zero-field splitting parameter. 
The figure \ref{fig:comparison-VV-C} shows the comparison between the Curie Law and the Van Vleck model.\\
In the range of temperatures reachable with a modern two-stages cryocooler ($3-10$ K), the difference in the two dc-susceptibility values is less than 20 percent. %Considering the current accuracy in the determination of the ionic concentration, we can claim that the Curie Law represents still well the temperature dependance of the dc-susceptibility. 

A second issue in the standard model arises from the case of Fe$^{3+}$ ion. The equation \ref{equ:chiprim}  is valid for a two levels system.
Fe$^{3+}$ has a spin $S=5/2$ and thus, its ground state is splitted in three Kramer's doublets: $|1/2\rangle, |3/2\rangle$ and $|5/2\rangle$. 
In the absence of a static magnetic field, there are thus three ESR at the frequencies  $12.0$ GHz, $19.3$ GHz and  $31.3$ GHz.
When dealing with a low power, the level populations stay almost those imposed by the thermal agitation. The susceptibility thermal behavior will be thus well 
represented by the Van Vleck model. At a high power, the differential saturation of these three transitions makes complexe the derivation of the equations. 
The $|1/2 \rangle \shortrightarrow |5/2\rangle$ transition at $31.3$ GHz is only allowed owing the state-mixing induced by the crystal field. For a frequency $\nu$ near 10 GHz, it is thus justified to neglect this transition when calculating  the evolution of overall magnetic susceptibility. About one third of the Fe$^{3+}$ ion population is on the  $|5/2\rangle$ level and is not affected by the RF magnetic field. We can reasonably conclude that the two levels model will lead to an overestimation of the impact of the ESR saturation on the magnetic susceptibility. We did not go further in the description of the Fe$^{3+}$ ion behavior as  the current uncertainties in the impurities concentration and in the ions relaxation times $\tau_{1}$ and $\tau_{2}$ make illusive 
a better quantitative analysis. 

%%%%%%%%%%%%%%%%%%%%%%%%%
%%%%%%%%%%%%%%%%%%%%%%%%%
\section{Whispering Gallery Mode Resonator: low power operation.}
\begin{figure}[h!]
\centering
\includegraphics[width=0.8\columnwidth]{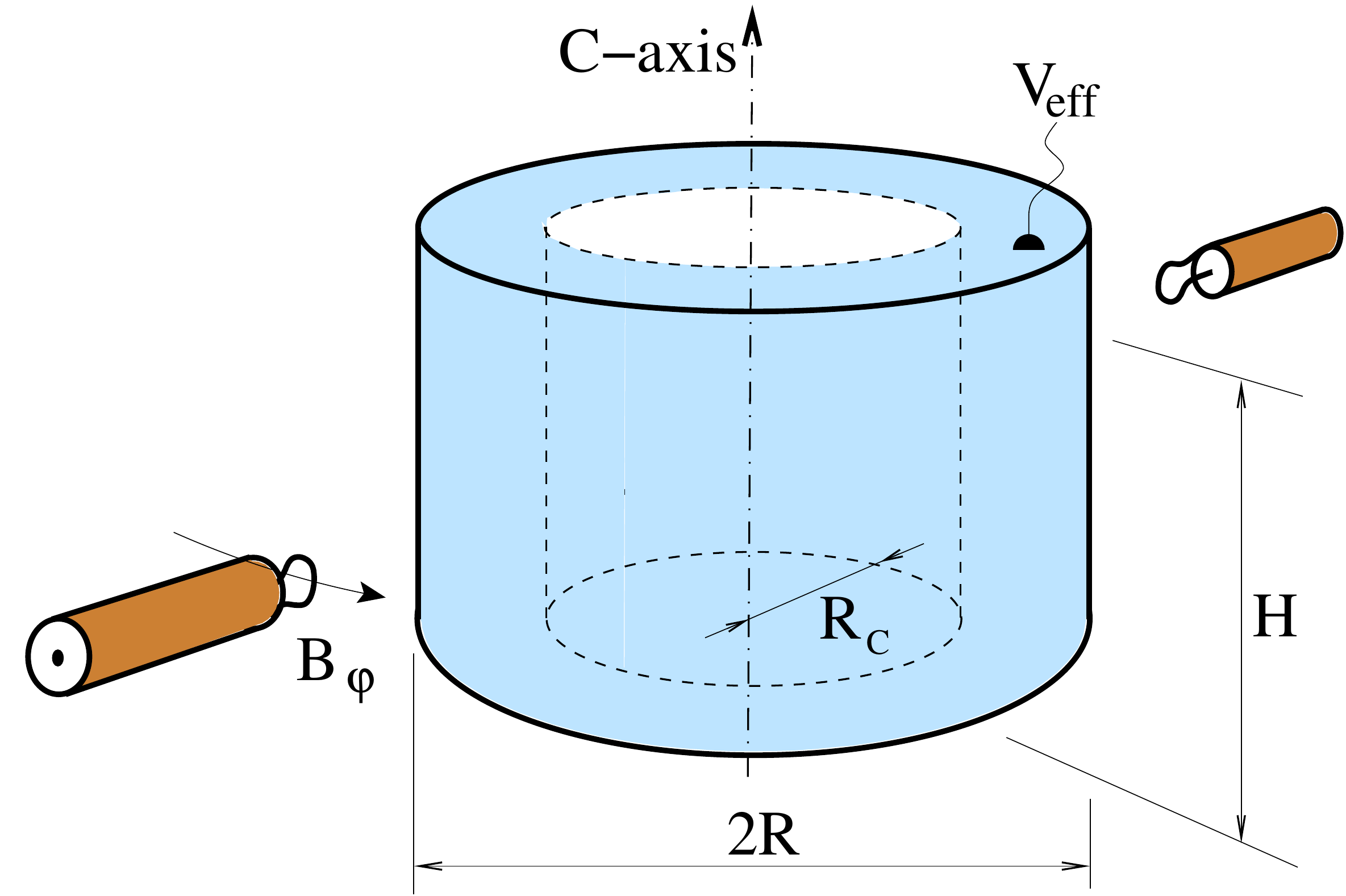}
\caption{The microwave whispering gallery mode resonator.}
\label{fig:waveguide}
\end{figure}

%%%%%%%%%%%%%%%%%%%%%%%%%
%%%%%%%%%%%%%%%%%%%%%%%%%
Due to its high-Q factor at low temperature the sapphire whispering gallery mode resonator constitutes a powerful tool to observe the behavior of the paramagnetic 
ions diluted in the crystal. The typical resonator geometry is shown in the figure \ref{fig:waveguide}.

 For an operation in X-band, the sapphire cylinder has a diameter $30 \leq 2R \leq55$ mm and a thickness $20 \leq H \leq30$ mm. 
 Its axis is parallel to the crystal C-axis within $\pm 0.5$ degree. The resonator is placed in the center of a cylindrical gold plated copper cavity and can be easily
 cooled down to about 3K inside a two-stages cryocooler developping a power of 0.5 W at 4K. Two diametrically opposed small magnetic loops excite the electromagnetic
 resonance modes of the structure. Among them, the high order hybrid modes called whispering gallery modes are characterized by a high degree of confinement of the 
 electromagnetic fields inside the dielectric medium.
In that case the resonator quality factor is only limited by the sapphire dielectric losses which are very low at acryogenic temperature \cite{braginsky87}. For a 10 GHz resonance mode 
the typical unloaded Q-factor is one billion at $6$K. For a whispering gallery mode, the energy density is confined between the cylindrical dielectric--air boundary 
and the inner caustic surface $\rho=R_C$\ (see Fig.  \ref{fig:waveguide}). Elsewhere the waves are evanescent. The resonator can thus be seen like a bent waveguide 
forming a ring \cite{mtt05_degenerescence}. The volume of this ring is $V_{eff}= \pi H (R^{2}-R_{C}^{2})$.
In the configuration shown in the figure \ref{fig:waveguide}, the magnetic field generated by the loops is perpendicular to the cylindrical axis allowing to excite the quasi-transverse magnetic whispering gallery modes such as $WGH_{m,n,l}$ mode. The three integers $m,n$ and $l$ represent the electromagnetic field components variations along the azimuthal $\varphi$, radial $\rho$ and axial $z$ directions respectively \cite{cros90}. We consider only the resonant modes with low radial and axial variations, i.e. those corresponding to $n = l = 0$ as they present the more efficient confinement inside the dielectric medium.

\begin{figure}[ht!]
	\centering
	\includegraphics[width=0.9\columnwidth]{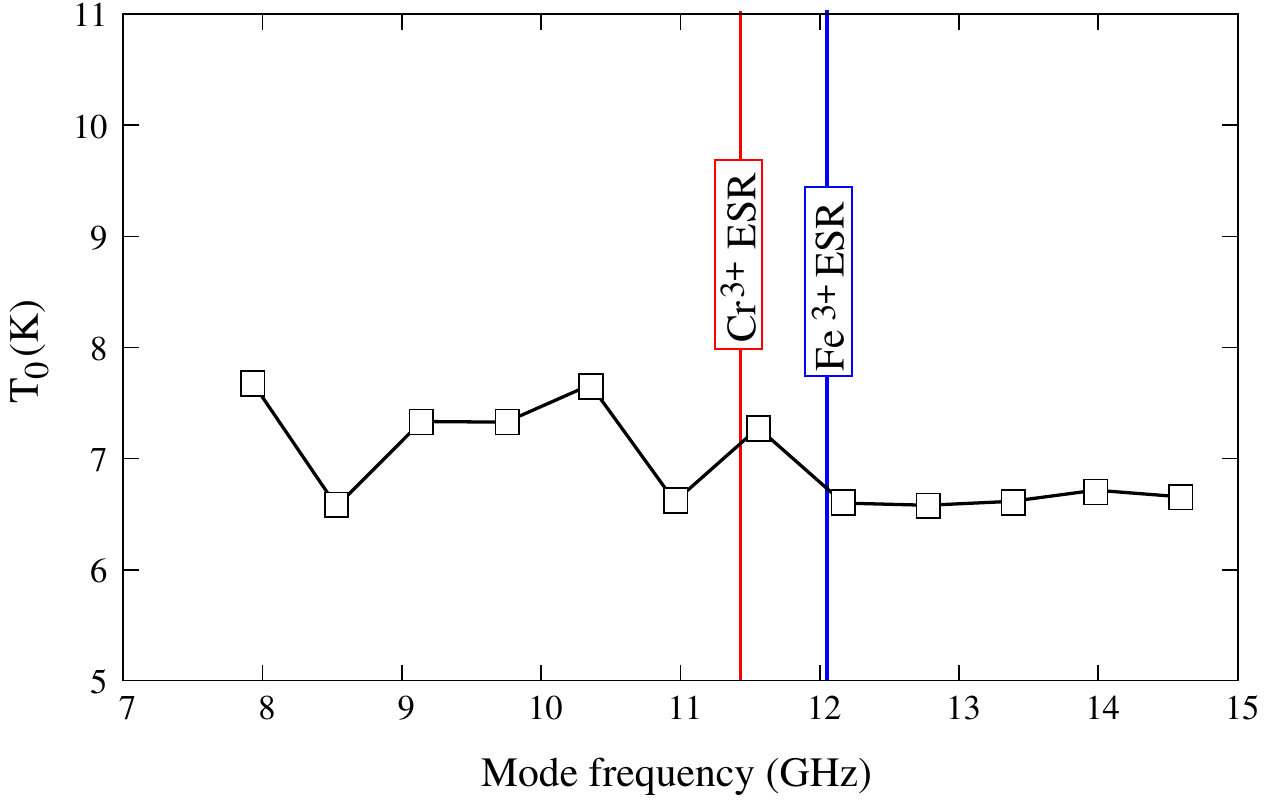}
	\caption{{$2R=50$ mm, $H=20$ mm HEMEX resonator: turnover temperature $T_{0}$ as a function of the mode frequency for $WGH_{m,0,0}$ modes with $4\leq m \leq 21$. }}
	\label{fig:inversion-les-3}
	\end{figure}

At low power, the temperature dependance of a given mode frequency $\nu$ is \cite{mann92-jpd}:
\begin{equation}
\dfrac{\nu(T)-\nu_{0}}{\nu_{0}}= A T^{4} - \eta \dfrac{\chi'(\nu,T)}{2}
\label{equ:fvsT}
\end{equation}
$\nu_{0}$ would be the mode frequency for a negligible circulating power, at $T=0$ K and in the absence of any paramagnetic dopant.
$A\approx  -3 \times 10^{-12}$ K$^{-4}$ \cite{luiten96} combines the temperature dependance of the dielectric constant and the thermal dilatation of the sapphire. 
The filling factor $\eta \approx 1$ for a high order whispering gallery mode. 
$\chi'$ is the real part of the ac susceptibility for a RF magnetic field perpendicular to the crystal C-axis. It is the sum of the contributions of  all ion species contained into the crystal. The low power assumption means that the thermal distribution of the population on the energy levels of the paramagnetic impurities is not modified by the RF magnetic field, i.e. $\Omega^{2}\tau_{1}\tau_{2} \ll 1$. For a mode frequency below the ESR of the dominant paramagnetic specie, the $1/T$ dependance of $\chi'$ will compensate for the intrinsic sapphire thermal sensitivity. The mode frequency passes through a maximum at a temperature $T_{0}$, which depends on the nature and concentration of the dopants. Our resonators are machined from HEMEX sapphire monocrystals provided by Crystal System Inc. \cite{schmid73}. Such a crystal is grown with the Heat Exchanger Method allowing the growth of large sapphire boule with the lowest defects and impurities concentration. The Fig. \ref{fig:inversion-les-3} shows the turnover temperatures as a function of the whispering gallery mode frequency for a $2R=50$ mm and $H=20$ mm sapphire resonator. 

\begin{figure}[h!!!!]
\centering
\includegraphics[width=0.95\columnwidth]{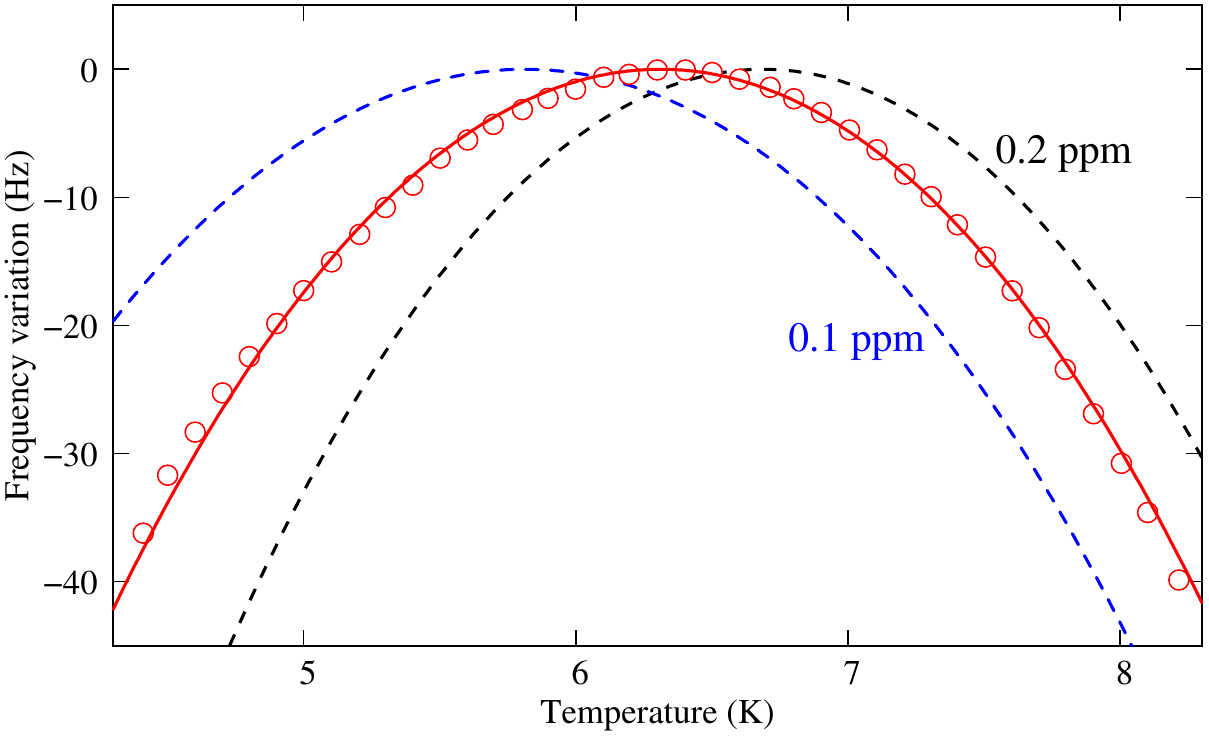}
\caption{WGH$_{16,0,0}$ mode frequency vs temperature variation for a 50 mm diameter and 20 mm hight HEMEX sapphire resonator. Red open circles: experimental data. 
Dashed lines: Frequency calculated with the Van Vleck
model with a Mo$^{3+}$ ions concentration of 0.1 ppm (blue dashed line), 0.2 ppm (black dashed line) and 0.15 ppm (solid red line).}
\label{fig:T0}
\end{figure}

All whispering gallery modes in a large frequency range present a turnover temperature almost independent of the mode order $m$. Luiten  \cite{luiten96} demonstrated that it is due to the predominance of the Mo$^{3+}$ ion, whose ESR frequency is 165 GHz. The Mo$^{3+}$ concentration was estimated to be of the order of  some $0.1$ ppm. The spread in turnover temperatures observed for low frequency modes ($\nu <12$ GHz) could result from Cr$^{3+}$ or/and Fe$^{3+}$ residuals. The concentration of these residuals should be very low as the turnover temperature imposed by the Mo$^{3+}$ ions is not greatly affected. Indeed other measurements show that  Cr$^{3+}$ and Fe$^{3+}$ concentrations are of some tens of ppb \cite{kovacich97,benmessai13-prb}. 
The figure \ref{fig:T0} shows the temperature dependance of the WGH$_{16,0,0}$ mode at 11.565 GHz of the same resonator and the theoretical predictions. The equation \ref{equ:fvsT} evaluated with $\chi_{0}^{VV}$ and $0.15$ ppm of Mo$^{3+}$ representes well the experimental frequency variation shown in the 
figure \ref{fig:T0}	

%%%%%%%%%%%%%%%%%%%%%%%%%
%%%%%%%%%%%%%%%%%%%%%%%%%
\section{Resonator power sensitivity}
%%%%%%%%%%%%%%%%%%%%%%%%%
%%%%%%%%%%%%%%%%%%%%%%%%%

\begin{figure}[h!!!!]
\centering
\includegraphics[width=0.85\columnwidth]{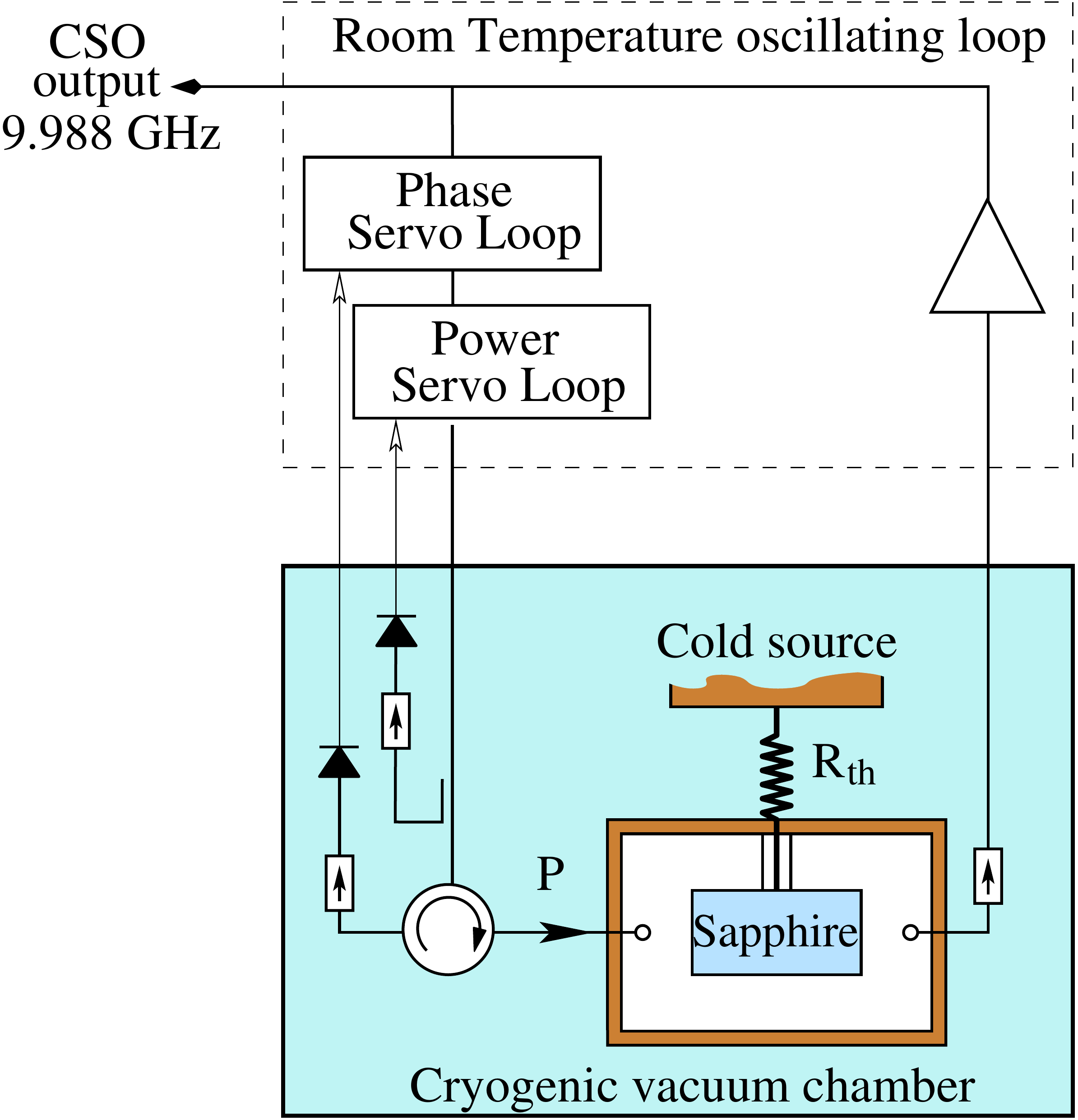}
\caption{Scheme of the Cryogenic Sapphire Oscillator. The cold source is a two-stages Pulse-Tube cryocooler.}
\label{fig:vg-oscillator-loop--mars2013}
\end{figure}

In the CSO the sapphire whispering gallery mode resonator is simply inserted in the positive feedback loop of an electronic amplifier to form an oscillator 
as schematised in the figure \ref{fig:vg-oscillator-loop--mars2013}. %
The CSO is completed by two servos to stabilize the power injected into the resonator and the phase lag along the sustaining loop \cite{rsi10-elisa}. The error signals needed for these two servos are derived from the low frequency voltages generated by two tunnel diodes placed near the resonator input port (see figure \ref{fig:vg-oscillator-loop--mars2013}). 
We build two identical oscillators: CSO-1 and CSO-2, a  third unit being under construction. These three instruments are intended to serve as references for ultra high resolution short term 
frequency stability measurements in the frame of the OSCILLATOR-IMP project \cite{www.osc-imp}.
The two CSOs are based on a $2R= 54$ mm and $H=30$ mm HEMEX resonator designed to operate on the quasi-transverse magnetic whispering gallery mode WGH$_{15,0,0}$ near $10$ GHz. For this resonator, the electromagnetic field is confined inside a volume $V_{eff} \approx 17$ cm$^{3}$.
The table \ref{tab:resonator-features} gives the current resonators characteristics as measured with a network analyzer using a $-10$ dBm probe signal.

\begin{figure}[!h]
\centering
\includegraphics[width=\columnwidth]{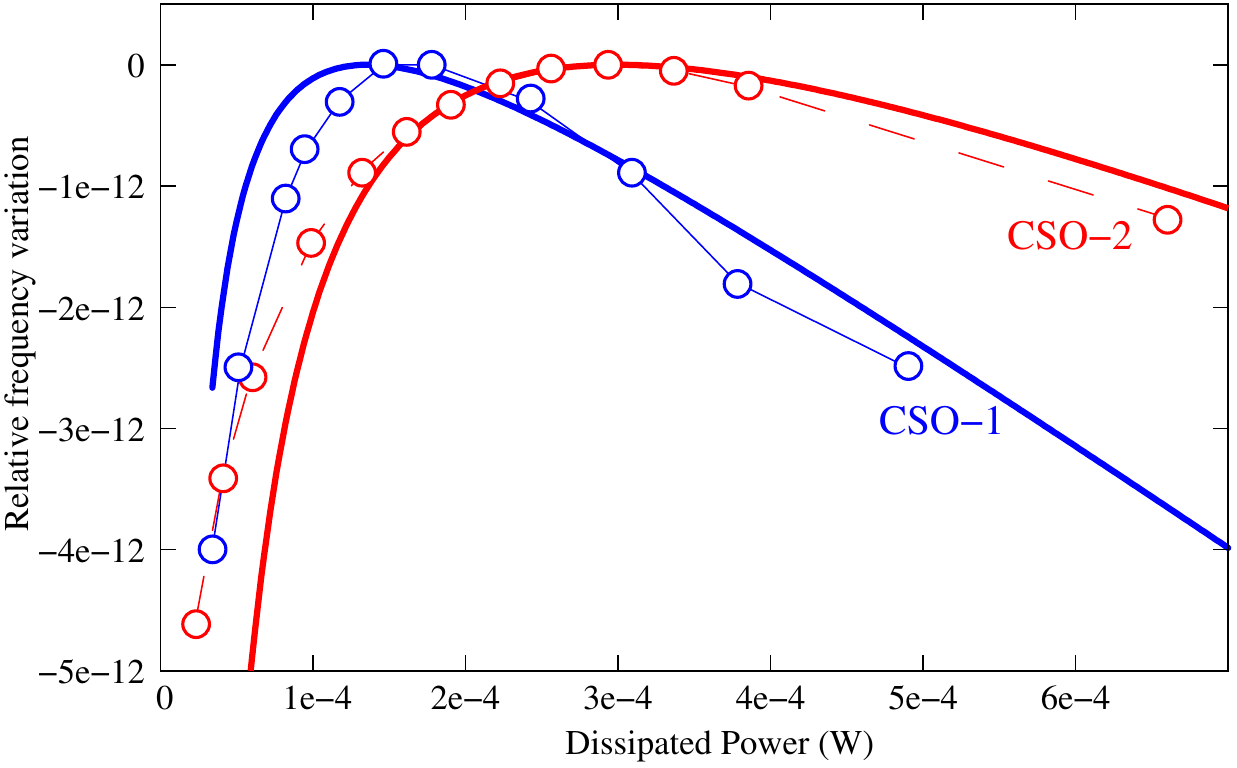}
\caption{Relative frequency variation vs dissipated power for COS-1 (blue) and CSO-2 (red). The open circles are the experimental points. The solid bold lines are the theoretical expectations calculated from the equation \ref{equ:fvsP} assuming a concentration of $0.2$ ppm of Mo$^{3+}$ and $10$ ppb of Cr$^{3+}$.}
\label{fig:power-dependance}
\end{figure}

\begin{table}[h!!!!]
\centering
\caption{The two resonators parameters: $T_{0}$: turnover temperature, $Q_{0}$: unloaded Q-factor, $\beta_{1}$ and $\beta_{2}$: coupling coefficient at the input and output ports respectively. $P_{0}$: turnover power at which the resonator frequency is maximum. $\frac{\Delta \nu}{\Delta P}$ the frequency power sensitivity slope  for $P \gg P_{0}$.}
\label{tab:resonator-features}
\begin{tabular}{lcccccc}
\hline
		&$T_{0}$	& $Q_{0}$ &  $\beta_{1}$ 	& $\beta_{2}$ & ~~~~~~$P_{0}$~~~~~~ &$ \frac{\Delta \nu}{\Delta P}$ at\\
		&		&	&	&	&		& high power\\
		& (K)		 &	&	&	&($\mu$W) & (Hz/W)\\
\hline
\hline	 
CSO-1 	& 6.23 	&~~~ $2.0\times 10^{9}$~~~	& 1 & 0.1 & 120 & -91 \\
CSO-2	& 6.18 	& $0.7\times 10^{9}$	& 1 & 0.1  & 300 &-39\\
\hline
\end{tabular}
\end{table} 
The Q-factor depends of the crystal quality but also of its cleaness. It can be affected by spurious modes and by some geometrical imperfections in the cavity symmetry or in the coupling probes alignement. Generally multiple cooldowns and fine step-by-step adjustments are required to get the highest unloaded Q-factor only limited by the sapphire dielectric losses. This was realized for CSO-1, still not for CSO-2 what explains its relative low Q-factor. 
For each resonator the coupling coefficients have been set near their optimal value, i.e.  $\beta_{1}\approx 1$ and $\beta_{2} \ll 1$. The injected power $P$ is almost entirely dissipated into the resonator. When the resonator is stabilized at its turnover temperature its thermal sensitivity nulls at first order and the CSO frequency stability is no longer limited by the cold source temperature fluctuations. The current limitation in the frequency instability is not clearly established. The resonator power to frequency conversion constitutes one possible limitation \cite{nand13} and needs thus to be investigated. To measure the CSO frequency sensitivity to the injected power, we follow the beatnote frequency changes when the power is varied in one CSO, while all other parameters being kept constant.  Tunnel diodes placed at a low temperature turn out to be very sensitive and can be dammaged if the incident power is too high. In the current resonator implementation the maximal injected power has been limited to about 1 mW. The figure \ref{fig:power-dependance} shows the relative frequency variation as a function of the injected power for CSO-1 and CSO-2.
The $WGH_{15,0,0}$ mode frequency passes through a maximum at a given power $P_{0}$ and exhibits for $P \gg P_{0}$  a linear negative sensitivity. The slope at high power $\frac{\Delta \nu}{\Delta P}$ and $P_{0}$, which depend on the resonator, are given in the table \ref{tab:resonator-features}.\\

 When the injected power $P$ is increased the resonator frequency will vary through different processes:
%
%
%%%%%%%%%%
\begin{description}
\item{\bf{Thermal effect:}\rm\ the resonator is linked by a thermal resistance $R_{th}$ to the cold source stabilized at the temperature $T_{S}$. The resonator temperature is $T=T_{S}+R_{th}P$. In the current design $R_{th}\approx 2$ KW$^{-1}$. Thus for $P= 1$ mW} the increase in the resonator temperature is only $2$ mK, which around the turnover temperature induces a relative frequency shift less than $5\times 10^{-14}$. This effect will give a negligible contribution in the experimentaly accessible range of power. 
\item{\bf{Radiation pressure:}\rm\ the stress induced by the stored energy results in a resonator expansion and a decrease in the dielectric constant. The resulting  resonator linear frequency to power sensitivity has been previously evaluated and can be written as \cite{chang97,nand13}:
\begin{equation}
\dfrac{\Delta \nu}{\Delta P}= \kappa\dfrac{Q_{0}}{V_{eff}}
\label{equ:radiation-pressure}
\end{equation}
with $\kappa\approx -7.2\times 10^{-13}$ Pa$^{-1}$.

In the preceeding works, the radiation pressure effect was assumed to be the major cause of the resonator power sensitivity. Indeed at a high power, this linear power dependance dominates the resonator frequency power sensibility. Equation \ref{equ:radiation-pressure} gives a sensitivity of $-85$ HzW$^{-1}$ and $-40$ HzW$^{-1}$ for CSO-1 and CSO-2 respectively, which are compatible with the experimental observations (see table \ref{tab:resonator-features}).
}
\item{\bf{ESR saturation:}\rm\ the third effect arises from the saturation of the ESR. To get an approximation of the RF magnetic field seen by the ions we neglect its space variations. Over the effective volume $V_{eff}$, we take it as a constant and equals to its mean value $\overline{B}$ defined as : 
\begin{equation}
\overline{B}^{2} = \dfrac{1}{V_{eff} }\iiint B B^{\star} dv 
\label{equ:E1}
\end{equation}
The stored energy is proportionnal to the power dissipated inside the resonator:
\begin{equation}
E_{stored} =\mu_{0} \overline{B}^{2} V_{eff}= \dfrac {Q_{0}P}{2\pi \nu_{0}}
\label{equ:E2}
\end{equation}
The amplitude of the ac-magnetic field can be thus written as:
\begin{equation}
\overline{B}= \sqrt{\dfrac {\mu_{0}Q_{0}P}{2\pi \nu_{0} V_{eff}}}
\end{equation}
With the typical resonator parameters, i.e. $Q_{0}=10^{9}$, $P=1$ mW, $\nu_{0}=10$ GHz and $V_{eff}=17$ cm$^{3}$, we find $\overline{B}= 1$ mT, 
which is about three orders of magnitude higher than the transverse magnetic field existing inside a 50 $\Omega$ coaxial cable where a 1 mW microwave signal is propagating. For a signal frequency $\nu$ not too far from the ESR frequency, i.e. $|\nu-\nu_{mn}| \leq \mathrm{few}\ \Delta \nu_{mn}$, the saturation of the ESR arises very rapidly when the injected power is increased. The ion energy level populations tend to balance themselves and the induced magnetic susceptibility goes to zero. For a signal power higher than $P_{0}$ only remains the linear power sensitivity imposed by the radiation pressure effect.
}\end{description}
The equation \ref{equ:fvsT} is now adapted to represent the resonator sensitivity to the injected power $P$:
\begin{equation}
\dfrac{\nu(P)-\nu_{0}}{\nu_{0}}= A (T_{S}+R_{th}P)^{4}+\dfrac{\kappa}{\nu_{0}}\dfrac{Q_{0}}{V_{eff}} P - \eta \dfrac{\chi'(\nu,P)}{2}
\label{equ:fvsP}
\end{equation}
This equation has been used to compute the relative frequency variation as a function of the dissipated power.
The result is given in the figure \ref{fig:power-dependance} (bold lines) assuming a concentration of 0.2 ppm of Mo$^{3+}$ and 10 ppb of Cr$^{3+}$ for both resonators. We found about the same shape in the power dependances by replacing chromium by 1 ppb of Fe$^{3+}$. As previously mentionned, the model overestimates the saturation of the ESR for Fe$^{3+}$, whose concentration has been measured of the order of 10 ppb in similar sapphire crystals. It should also be pointed out that the value of $P_{0}$ is greatly dependant of the relaxation times $\tau_{1}$ and $\tau_{2}$, which are not known with accuracy. Nevertheless our model explains qualitatively well the power sensitivity of the sapphire resonator.

%%%%%%%%%%%%%%%%%%%%%%%%%
\section{Application to the realization of an ultra stable oscillator}
\begin{figure}[!h]
\centering
\includegraphics[width=\columnwidth]{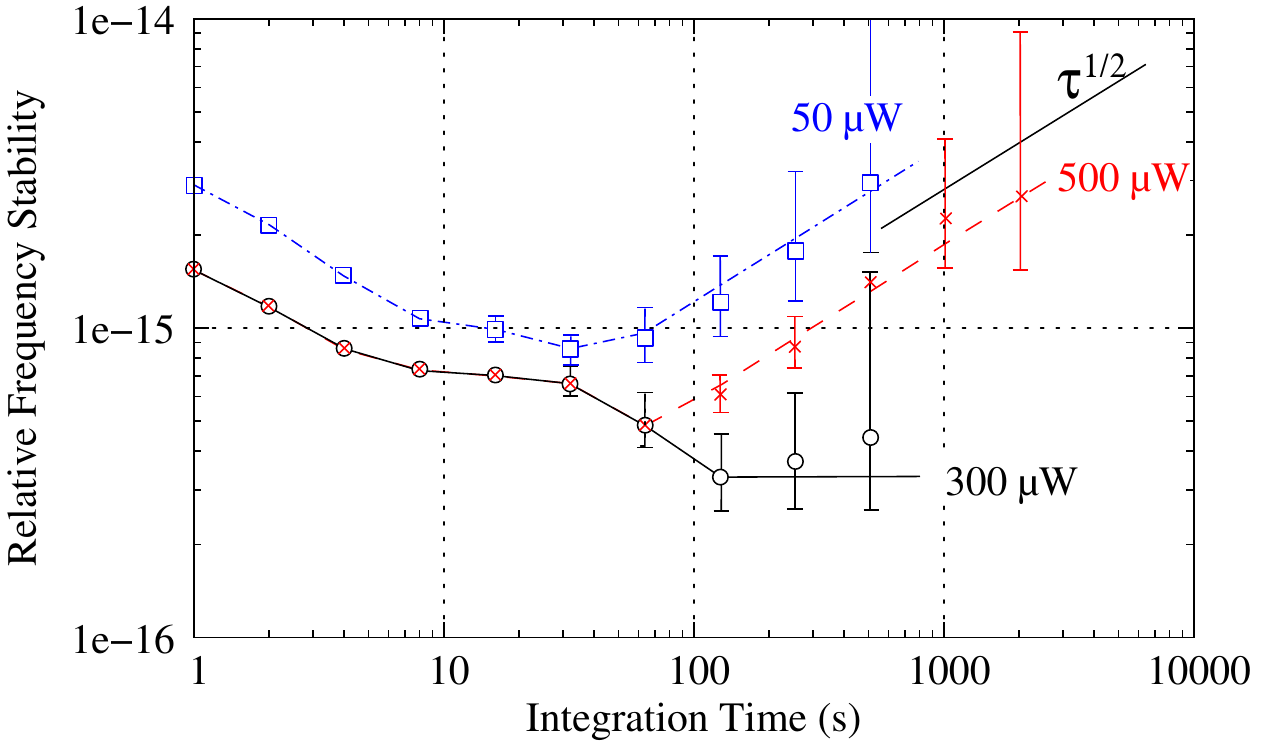}
\caption{Measured frequency stability (Allan standard deviation) for different values of the power injected in CSO-2. CSO-2 operated in a degraded mode without power control. }
\label{fig:stab-vs-injected-power-v2}
\end{figure}

%%%%%%%%%%%%%%%%%%%%%%%%%
We can conclude from the previous observations (see figure \ref{fig:power-dependance}) that it exists for each resonator a value of the injected power, i.e. $P_{0}$ for which the sensitivity to the power fluctuations nulls to the first order. At that point the CSO frequency stability would not any more be limited by the fluctuations of the injected power. To verify this assumption, we conducted the following measurement: CSO-2 was operated in a degraded mode with its power servo in open loop. A laboratory DC-power supply was used to bias the voltage controlled attenuator (VCA) placed in the oscillator loop and that controls the level of the injected power. The DC voltage $V$ generated by the power supply fluctuates with time following a random walk process. We observed that its standard deviation $\sigma_{V}(\tau)$ when averaged over $\tau =1$ s is of the order of  $10$  $\mu$V.  At longer integration time, i.e. $\tau \geq 10$ s,   $\sigma_{V}(\tau)$ is degraded proportionally to $\tau^{1/2}$, typical of a random walk process. Through the VCA the power injected in the resonator and thus the CSO frequency are modulated by this voltage noise. The figure \ref{fig:stab-vs-injected-power-v2} shows the relative frequency stability (Allan standard deviation) mesured by beating the two CSOs for different values of the power injected in CSO-2 imposed by the DC-voltage $V$. CSO-1 was nominally running with its power servo on.

When the injected power is tuned to $P_{0}=300\ \mu$W, the gain of the phase servo is optimal and the measured short term frequency stability at $\tau=1$ s is $1.5\times 10^{-15}$, which includes the contributions of both CSOs. At this level of power the CSO frequency stability would not be limited by the fluctuations of the injected power.
Indeed, the relative frequency stability improves as $\tau$ is increased to reach a flicker floor of  $3\times 10^{-16}$. 
At a high power, i.e. $P=500\ \mu$W, the short frequency stability remains unaltered. Then for $\tau \geq100$ s, the frequency stability is clearly limited by random walk process. The measured frequency noise level is compatible with the DC-voltage noise and the resonator sensitivity at such a high injected power.
At low injected power, $P=50\ \mu$W, the short term frequency stability for $\tau \leq10$ s is degraded as the gain of the phase servo, which is proportional to $P$, is decreased. At longer integration time $\tau \geq 100$s, the measured frequency stability is degraded proportionally to $\tau^{1/2}$ with a frequency noise level higher than for $P=500 \mu$W as the power sensitivity at low injected power is higher.

Eventually, in both oscillators, the resonator temperature and the injected power were stabilized at their inversion point, i.e. $T_{0}$ and $P_{0}$ respectively. The relative frequency stability is measured by beating the signals of the two identical CSOs, separated by 7.029 MHz, with a frequency counter without dead time. 
The next figure shows the raw data recorded for a quiet period of four hours.

\begin{figure}[h!!!!!]
\centering
\includegraphics[width=0.94\columnwidth]{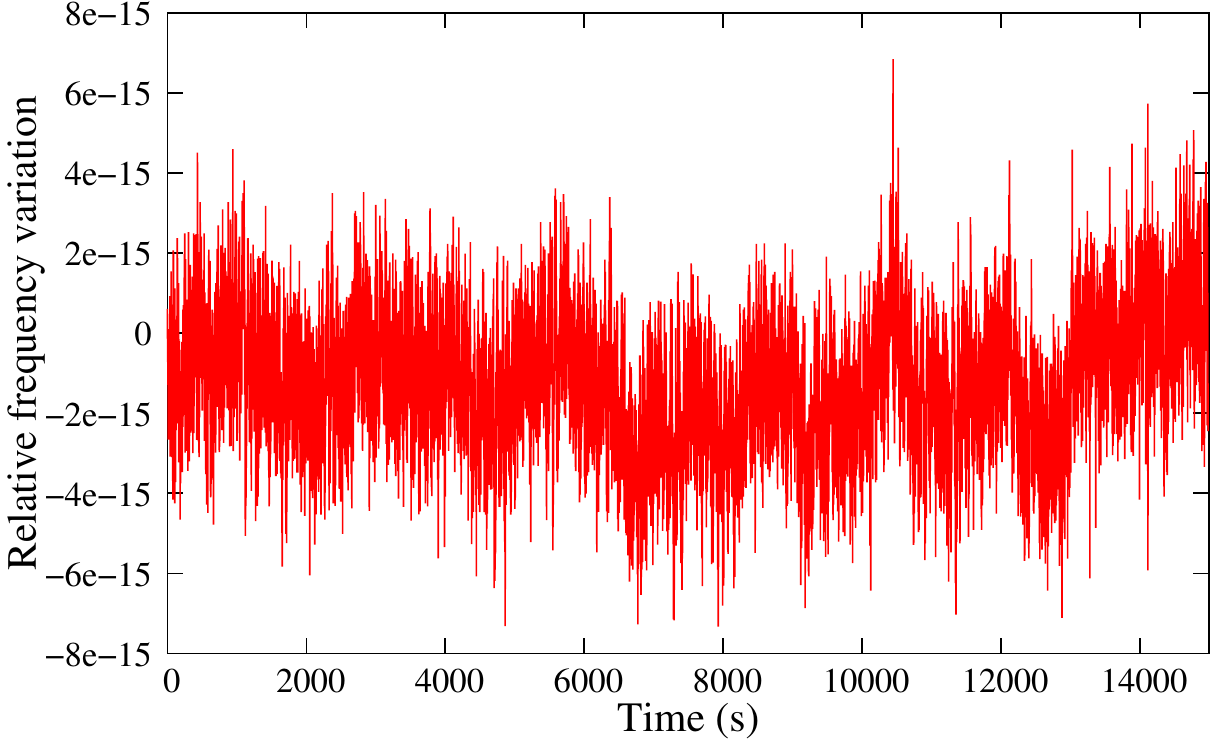}
\caption{Relative frequency variation during a periode of about four hours measured by beating two almost identical optimized CSOs.}
\label{fig:data-131210_131212}
\end{figure}

\begin{figure}[!h]
\centering
\includegraphics[width=\columnwidth]{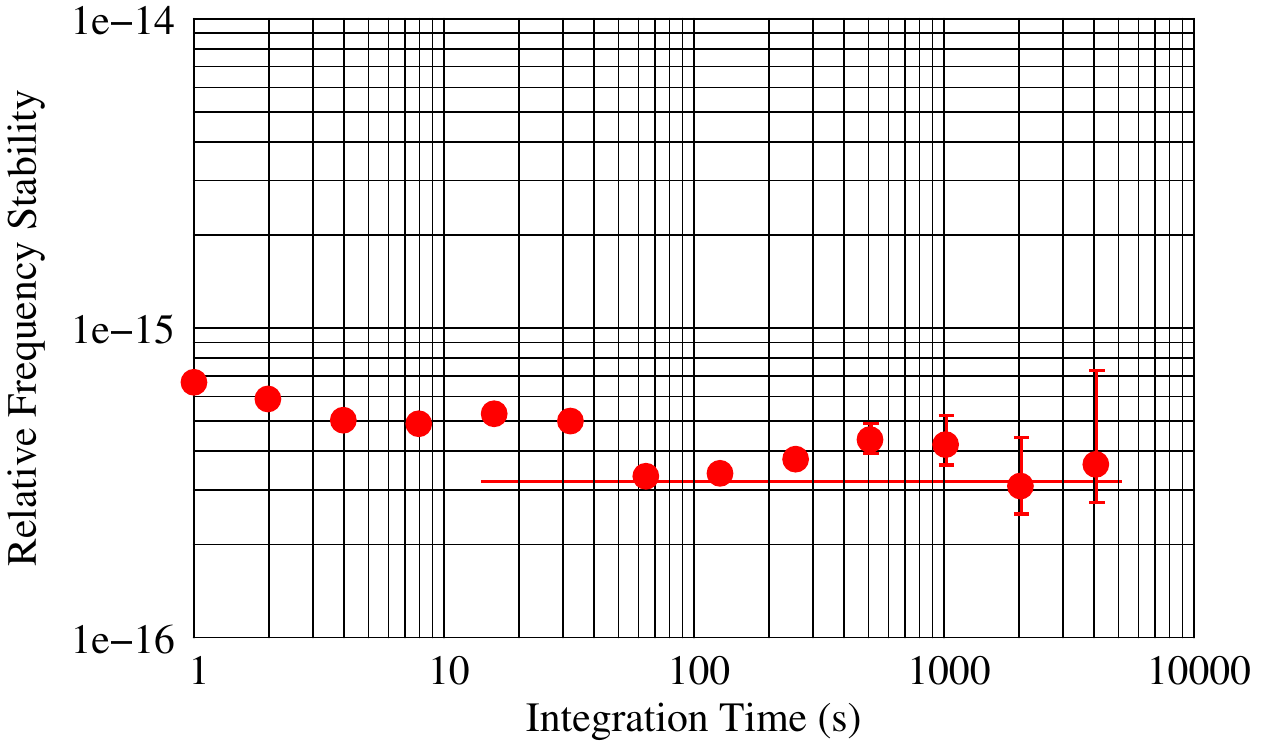}
\caption{Relative frequency stability of one optimized CSO. The Allan deviation has been calculated from the raw data of the figure \ref{fig:data-131210_131212} and thanks to the freeware \it{SigmaTheta}\rm\ \cite{vernotte2012} available on \cite{www-sigmatheta}
.}
\label{fig:best-stab-2014-01}
\end{figure}

The relative frequency stability (Allan standard deviation) of one CSO is presented in the  figure \ref{fig:best-stab-2014-01}: 3 dB was substracted considering that the total measured noise is  the sum of the contribution of two identical oscillators.\\

For $1$ s $\leq \tau \leq 10,000$ s the relative frequency stability is better than $7\times 10^{-16}$. A flicker floor of  3$\times 10^{-16}$ is reached at 100 s. The first small hump appearing around $\tau=30$ s could be due to a residual pumping we have detected in the temperature servo of one resonator. The second hump around $\tau= 800$ s comes from the pumping of the laboratory climatisation. This second hump could be due to the frequency stability measurement instrument sensitivity to the room temperature.

%%%%%%%%%%%%%%%%%%%%%%%%%
\section{Conclusion}
%%%%%%%%%%%%%%%%%%%%%%%%%
In summary we exploited the Van Vleck model to describe the saturation of the electron spin resonance of the paramagentic species contained in a high quality sapphire resonator. The proposed model explains qualitatively well the frequency-to-power sensitivity of the sapphire resonator. The ESR saturation and the sapphire intrinsic sensitivity compensate themselves at a given injected power, leading to a turnover in the frequency-vs-power curve. It is thus demonstrated than the CSO stabilized at this power value presents an exceptional short term frequency stability better than  $7\times 10^{-16}$ for $\tau \leq 10,000$ s.

%%%%%%%%%%%%%%%%%%%%%%%%%
\section{Acknowledgements}
%%%%%%%%%%%%%%%%%%%%%%%%%
The work has been realized in the frame of the ANR projects: Equipex Oscillator-Imp and Emergence ULISS-2G. The authors would like to thank the Council of the R\'egion de Franche-Comt\'e for its support to the \it{Projets d'Investissements d'Avenir}\rm\ and the FEDER for funding one CSO.\rm

\end{document}